\documentclass[twocolumn,amsmath,amssymb]{snp}
\usepackage{graphicx}% Include figure files
\usepackage{epsfig}
\usepackage{dcolumn}% Align table columns on decimal point
\usepackage{bm}% bold math
\usepackage{color}
\def\J{$J/\psi$}

\def\U{$\Upsilon$}

\newcommand{\jpsi}{J / \psi}

\newcommand{\old}[1]{}

\newcommand{\be}{\begin{equation}}
\newcommand{\ee}{\end{equation}}
\newcommand{\ba}{\begin{eqnarray}}
\newcommand{\ea}{\end{eqnarray}}
\newcommand{\bi}{\begin{itemize}}
\newcommand{\ei}{\end{itemize}}

\begin{document}
\title{{\Large  $J/\psi$ suppression in nucleus-nucleus collisions
}}% Force line breaks with \\
\author{\large $^*$Binoy Krishna Patra$^1$}
\author{\large Vinod Chandra$^2$}
\author{\large Vineet Agotiya$^1$}
\affiliation{$^1$ Department of Physics, Indian Institute of 
Technology Roorkee, India, 247 667}
\affiliation{$^2$ Department of Physics, Indian Institute of Technology Kanpur,
India, 208 016 }
\begin{abstract}
\leftskip1.0cm
\rightskip1.0cm
At high temperatures, strongly interacting matter becomes a plasma of
deconfined quarks and gluons. In statistical QCD, deconfinement and the
properties of the resulting quark-gluon plasma can be investigated by
studying the in-medium behaviour of heavy quark bound states.
In high energy nuclear interactions, quarkonia probe different aspects of the
medium formed in the collision. So, we first reviewed the fate of
quarkonia in the different stages of the (dynamical) system produced at the
collision. We have then presented our present work on the
dissociation of the heavy quarkonium states in a hot QCD medium by 
investigating the medium modifications to heavy quark 
potential. In contrast to the usual screening picture, interestingly 
our theory gives rise the screening of the charge, not the range of the 
potential.
\end{abstract}
\maketitle
\noindent
{\large{\bf Introduction:}}
The study of the fundamental forces between quarks and gluons is an essential
key to the understanding of QCD and the occurrence of different phases which
are expected to show up when going from low to high temperatures ($T$) and/or
baryon number densities. For instance, at small or vanishing
temperatures quarks and gluons get confined by the strong force while at
high temperatures asymptotic freedom suggests a quite different QCD medium
consisting of rather weakly coupled deconfined quarks and gluons, the 
so-called quark gluon plasma (QGP).
The anomalous suppression of the $J/\psi$ production in heavy ion collisions
which has been experimentally observed~\cite{Abr00} in the depletion 
of the dilepton multiplicity in the region of invariant mass corresponding 
to the $J/\psi$ meson was proposed long time ago as a possibly unambiguous 
signal of the onset of deconfinement. Matsui and Satz~\cite{Mat86} 
argued that charmonium states produced before the formation of a 
thermalized QGP would tend to melt in their path 
through the deconfined medium, since the 
binding (colour) Coulomb potential is screened by the large number of colour 
charges. This, in turn, would produce an anomalous (with respect to normal 
nuclear absorption) drop in the $J/\psi$ yields.

In this picture it is implicitly assumed that, once the charmonium dissociates,
the heavy quarks hadronize by combining with light quarks only
(recombination leading to a secondary $J/\psi$ production is neglected). This
assumption is certainly justified at the SPS conditions, due to the very small
number of $c\bar{c}$ pairs produced per collision ($N_{c\bar{c}}\sim 0.2$ in a
central collision), but at RHIC ($N_{c\bar{c}}\sim 10$) and LHC
($N_{c\bar{c}}\sim 200$) energies it is no longer warranted~\cite{raf}.

Moreover in a hadronic collisions only about $60\%$ of the observed $J/\psi$'s
are directly produced, the remaining stemming from the decays of excited
charmonium states (notably the $\chi_c$ and the $\psi'$). Since each $c\bar{c}$
bound state dissociates at a different temperature, a model of {\em sequential
suppression} was developed, with the aim of reproducing the $J/\psi$
suppression pattern as a function of the energy density reached in the heavy
ion collision. SPS experimental data for Pb-Pb collisions at different 
centralities seem indeed to support the dissociation pattern predicted by 
this model.\\

The heavy quark pair leading to the $J/\psi$ mesons are produced in 
nucleus-nucleus collisions on a very short time-scale
$\sim 1/2m_{c}$, where $m_c$ is the mass of the charm quark. The pair
develops into the physical resonance over a formation time $\tau_\psi$
and traverses the plasma and (later) the hadronic matter
before leaving the interacting system to decay (into  a dilepton) to
 be detected. This long `trek' inside the interacting system is fairly
`hazardous' for the $J/\psi$. Even before the resonance is formed
it may be absorbed by the nucleons streaming past it~\cite{GH}. By the
time the resonance is formed, the screening of the colour forces
in the plasma may be sufficient to inhibit a binding of the $c\overline{c}$
~\cite{Mat86}. Or an energetic gluon~\cite{xu} or a comoving
hadron could dissociate the resonance(s). 

Quarkonia at finite temperature are an important tool for the study of
quark-gluon plasma formation in heavy ion collisions. 
Many efforts have been devoted to determine the 
dissociation temperatures of $Q\bar{Q}$ states in the deconfined 
medium, using either lattice calculations of quarkonium spectral functions
or non-relativistic calculations based upon some effective (screened)
 potential.

Lattice studies are directly based on quantum chromodynamics and should
provide, in principle, a definite answer to the problem.
However, in lattice studies the spectral functions have to be extracted ---
using rather limited sets of data --- from the Euclidean (imaginary time)
correlators, which are directly measured on the lattice.
This fact, together with the intrinsic technical difficulties of lattice
calculations, somehow limits the reliability of the results obtained so far,
and also their scope, which in fact is essentially limited to the 
mass of the ground state in each $Q\bar{Q}$ channel.  
Potential models, on the other hand, provide a simple and intuitive framework
for the study of quarkonium properties at finite temperature, allowing one to
calculate quantities that are beyond the present possibilities for lattice
studies. The main problem  of the latter approach is the determination of the
effective potential: although at zero temperature the use of effective
potentials and their connection to the underlying field theory is well
established, at finite $T$ the issue is still open.

Calculations of the $c\bar{c}$ and $b\bar{b}$ dissociation temperatures, using
different potential models based upon the lattice free and internal energies,
have found on the whole a reasonable agreement with the results from the
lattice studies~\cite{Won07,Alb07}. On the other hand, calculations of 
Euclidean correlators using a variety of potential models were
not able to reproduce the temperature dependence of the lattice correlators. 

A precise quantitative agreement with the lattice correlators should not be 
expected, because of uncertainties coming from a variety of sources. Not 
only the determination of the effective potential is still an open question but
there are also issues tied, e.g., to relativistic effects, to the
thermal width of the states or to the contribution of radiative 
corrections. On the other hand, lattice correlators are
also affected by their own uncertainties. These may be due
to the use of different lattices (isotropic or anisotropic); to
the finite size of the box, which might significantly alter the
continuum part of the spectrum, although calculations with
boxes of different sizes show discrepancies below 1\%
[5,6]; or to artifacts in the continuum region of the spectral
functions due to the finite lattice spacing [5].

Recently, Umeda and Alberico~\cite{Ume07,Alb08} have shown that the 
lattice calculations of meson correlators at finite temperature 
contain a constant contribution, due to the presence of zero modes 
in the spectral functions. These contributions cure most of the 
previously observed discrepancies with lattice calculations, 
supporting the use of potential models at finite temperature 
as an important tool to complement lattice studies.

Actually, even if the potential supports the existence of bound states, other
physical processes may lead to the dissociation of the quarkonium. First, if
the $Q\overline{Q}$ binding energy is lower than the temperature --- and
assuming that the quarkonia have reached the thermal equilibrium with the
plasma --- a certain fraction of their total number will be thermally excited
to resonant states according to a Bose-Einstein distribution: such a process is
referred to as {\em thermal dissociation}. Furthermore, the collisions 
with the gluons and the light quarks of the plasma
may lead to the {\em collisional dissociation} of the quarkonium~\cite{xu}. 

Binoy and Menon revisited the $J/\psi$ suppression due to 
gluonic bombardment in an expanding quark-gluon plasma in
a series of works~\cite{g-psi}. First they neatly incorporated 
the crucial effects arising from gluon fugacity, relative $g-\psi$ flux, 
and \J meson formation time and then used these effects in the formulation 
of the gluon number density, velocity-weighted cross section, and the 
survival probability in an equilibrated static QGP. This formulation have 
been used to study the pattern of $J/\psi$ suppression in the 
central rapidity region at RHIC/LHC energies. 
Later they explicitly take into account the effect of hydrodynamic 
(longitudinal) expansion profile on the gluonic breakup 
of $J/\psi$'s in an (chemically) equilibrating expanding QGP. 
A novel type of partial-wave {\em interference} 
mechanism is found to operate in the modified dissociation rate.
Finally, this formulation has been applied to the case 
when the medium is undergoing cylindrically symmetric {\em transverse} 
expansion. Compared to the case of longitudinal expansion the new graph of
survival probability develops a rich structure at RHIC, due to
a competition between the transverse catch-up time and plasma 
lifetime.

Of course, if the process $g+J/\psi\rightarrow c+\overline{c}$,
discussed above, can lead to the
dissociation of the charmonium, the same reaction can also occur in the
opposite direction. Hence a consistent calculation of \J  multiplicity
implies the solution of a kinetic rate equation integrated over the lifetime of
the QGP phase in which both processes (dissociation and recombination) enter
~\cite{raf} This is of relevance because, as mentioned above, the
usual assumption in considering the \J suppression as a signature of
deconfinement is that its production can occur only in the very initial stage
of the collision. Really, if at SPS the role played by recombination is
numerically negligible, this is no longer true at RHIC as pointed out in
Ref.~\cite{raf}.

However, there is a generic question, quite often asked, is  
that whether we can distinguish between the two mechanisms of dissociation 
(colour screening and collision with hard gluons) mainly operating in the
deconfined phase. Binoy and Srivastava~\cite{dks} have shown that 
while the gluonic dissociation of the \J is always possible, the Debye 
screening is not effective in the case of small systems at RHIC energies.
For the larger systems, the Debye screening
is more effective for lower transverse momenta, while the gluonic
dissociation dominates for larger transverse momenta. At LHC energies
the Debye screening is the dominant mechanism of $J/\psi$ suppression
for all the cases and momenta studied.
As an interesting result, they found the gluonic dissociation to be 
substantial but the Debye screening to be ineffective for $\Upsilon$ 
suppression at the LHC energy.

In the context of \J  suppression, Langevin dynamics seems to be 
almost as important as other mechanisms invoked to explain the RHIC and LHC 
data. Binoy and Menon~\cite{lang} considered the Brownian motion of 
a $c \bar c$ pair produced in the very early stage of a QGP. 
They found that, in the weak coupling regime, both the time 
scales associated with the positional swelling ($\tau_x$) and 
approach to ionization ($\tau_E$) are positive and less 
than the frictional relaxation time ($\gamma^{-1}$). Hence Brownian movement 
can cause a genuine break-up of the $c \bar c$ bound by swelling it 
substantially or Langevin dynamics can cause $ c \bar c$ to ionize after a 
time span $\tau_E$. On the other hand, in the strong coupling case, 
$\tau_x$ is imaginary and $\tau_E$ are negative, i.e., unphysical. 
Hence random force plus ̄diffusion cannot cause the $c \bar c$ bound state 
to dissociate.

While the short and intermediate distance ($rT \le1$) properties
of the heavy quark interaction is important for the understanding
of in-medium modifications of heavy quark bound states, 
the large distance property of the heavy quark interaction which 
is important for our understanding of the bulk properties of
the QCD plasma phase {\em viz.} the equation of state. 
In all of these studies deviations from perturbative calculations and the ideal gas
behaviour are expected and were indeed found at temperatures which are only
moderately larger than the deconfinement temperature. This calls for
quantitative non-perturbative calculations. 
The phase transition in full QCD will appear as an crossover
rather than a 'true' phase transition with related singularities in
thermodynamic observables (in the high-temperature and low density regime) 
a cross-over, it can be reasonable to assume that
the string-tension does not vanish abruptly above $T_c$. So we~\cite{vinet}
decide to investigate in our present work what happens to the different 
quarkonium states
if one corrects with a dielectric function encoding the effects of the
deconfined medium,the full Cornell potential and not only its Coulomb
part as usually done in the literature. We found that with this
choice medium effects give rise to a long-range Coulomb potential with a
reduced effective charge (inversely proportional to the square of the 
Debye mass) of the heavy quark, at variance with its usual
Debye-screened form employed in most of the literature. With such an
effective potential we investigate the effects of different
possible choices of the Debye mass on the dissociation temperature of
the different quarkonium states. Since a Coulomb interaction always
admits bound states, a criterion has to be adopted to define such a
dissociation temperature: a state is then considered ``melted" when its
binding energy becomes of the same order as the temperature.\\

\noindent
{\large{\bf The Debye masses in hot QCD:}}

Adequate knowledge of  Debye mass is indeed needed to study 
the medium modifications to heavy quark potential. 
The Debye mass in QCD unlike QED is generically non-perturbative and 
gauge invariant\cite{arnold}. The Debye mass at high temperature in the 
leading-order in QCD coupling is known from long time and is 
perturbative~\cite{shur}. The Debye mass in leading-order from the 
polarization tensor of a gauge boson derived from the HTL approach can 
also be obtained from the transport theory~\cite{trans}. One 
cannot naively generalize the definition of Debye mass in QED 
to QCD due to the non-abelian nature of the theory. Rebhan~\cite{rebh} has 
defined Debye mass through the relevant pole of the static quark 
propagator instead of the zero momentum limit of the time-time component of 
the gluon self-energy. The Debye mass thus defined comes out to be gauge 
independent follows from the fact that the pole of the self-energy is 
independent of choice of gauge. Braaten and Nieto~\cite{braaten} 
computed the Debye screening mass for QGP at high temperature to the 
next-to-leading-order in QCD coupling from the correlator of two 
Polyakov loops which agreed to the HTL result of~\cite{rebh}.

Arnold and Yaffe~\cite{arnold} pointed out that
the contribution of order $(g^2T)$ to the Debye mass in QCD  needs the 
knowledge of the non-perturbative physics of confinement of magnetic 
charges and a perturbative definition of the Debye mass as a pole of gluon 
propagator no longer holds. They showed how one can define Debye 
mass in QCD in a manifestly gauge invariant manner (in vector-like gauge 
theories with zero chemical potential). 
In the work of Kajantie {\it et. al} 
~\cite{kaj1}, the non-perturbative contributions of $O(g^2T)$ and 
$O(g^3T)$ have been determined from 3-D effective field theory which 
we consider in the present work. At high temperatures and
zero chemical potential Debye mass can be expanded in a power
series in QCD coupling~\cite{kaj1}:
\begin{eqnarray}
\label{eq}
m_D & = & m_D^{LO}+{Ng^2T\over4\pi}\ln{m_D^{LO}\over  
g^2T}\nonumber\\
 & + & 
c_{{}_N} g^2T + d_{{}_{N,N_f}} g^3 T + {\cal O}(g^4T) ~,
\end{eqnarray}
where $m_D^{LO}$ is the leading-order result~\cite{shur}.
The coefficient $d_{N,N_f}$ have the following dependence on the number 
of colors $N$ and flavors $N_f$ as:
\begin{eqnarray}
\label{eq1}
d_{N,N_f} & = &  \frac{b_N }{\sqrt{N/3+N_f/6}} \quad,
\end{eqnarray}
where the values of $c_{{}_N}$, $b_N$ have been obtained by fitting the results
with the physical 4D finite temperature QCD~\cite{kaj1}:
\begin{eqnarray}
{\rm{SU(3):}} c_{{}_N} = 2.46\pm 0.15\,\, b_N =-0.49\pm 0.15
\end{eqnarray}
The number $c_{{}_N}$ captures the  non-perturbative 3-D effects, while the 
$d_{N,N_f}$ is related to the choice of scale in $m_D^{\rm LO}$. We employ 
the two-loop expression for the QCD coupling constant at finite 
temperature. We use the following notations henceforth,
\ba
\label{notation}
m^{LO}_D &= & g(T) T \sqrt{\frac{N}{3}+\frac{N_f}{6}} \nonumber\\
m^{NLO}_D&=&m_D^{\rm{LO}}+{Ng^2T\over4\pi}\ln{ m_D^{\rm{LO}}\over  
g^2T}\nonumber\\
m^{NP}_D&=& m^{NLO}_D +c_{{}_N} g^2T + d_{N,N_f} g^3 T\nonumber\\ 
m^{L}_D&=&1.4 m^{LO}_D \quad,
\ea
where $m^{L}_D$ is the Debye mass obtained by fitting the (colour-singlet)
free energy in lattice QCD~\cite{dm_lattice}. 

In the weak coupling ($g<<1$) regime, the soft scale ($m_D \simeq gT$ 
at the leading-order) related to the screening of electrostatic fields 
is well separated from the ultra-soft scale ($\simeq g^2T$) 
related to the screening of magnetostatic fields. In such regime
it appears meaningful to see the contribution of each terms
in the the Debye mass (Eq.~{\ref{eq}) separately. But when the 
coupling becomes large enough (which is indeed the case), the 
two scales are no longer 
well separated. So while looking for the next-to-leading corrections 
to the leading-order result from the ultra-soft scale, it is not a wise
idea to stop at the logarithmic term (as mentioned in the notation 
$m^{NLO}_D$), since it becomes crucial the number multiplying the 
factor $1/g$ to establish the correction to the LO result. 
In fact we found that the Debye mass in the NLO term
($m_D^{\rm{NLO}}$) is always smaller than than the LO term ($m_D^{\rm{LO}}$) 
because of the negative (logarithmic) contribution ($\log (1/g)$)
to the leading-order term, while the full correction (all $g^2T$ terms)
to the Debye mass results positive. So, we will employ only 
three forms of the 
Debye masses {\em viz.} leading-order term/perturbative result
($m_D^{\rm{LO}}$), full (non-perturbative) corrections to the
leading-order term ($m_D^{\rm{NP}}$), and lattice parametrized 
form ($m_D^L$) to study the dissociation phenomena of quarkonium 
in a hot QCD medium in this work.
 
We now proceed to  investigate
in-medium modifications to heavy-quark potential and its application
to determine the binding energy and dissociation temperature
of the heavy-quark bound states.\\

\noindent
{\large{\bf The in-medium heavy-quark potential:}}

Let us now turn our attention to study the
medium modifications to heavy quark potential at $T=0$
which is considered as the Cornell potential,
\begin{equation}
\label{eqc}
 V(r)=-\frac{\alpha}{r}+\sigma r 
\end{equation}
where $\alpha$ and $\sigma$ are the phenomenological parameters. The former 
accounts for the effective coupling between the heavy quark pairs and the 
latter gives the string coupling.The medium modifications enters in the 
Fourier transform of the heavy quark potential as follows:
\begin{equation}
\label{eq3}
%\tilde {\bf V}(k)=\frac{{\matcal V}(k)}{\epsilon(k)}
\tilde {V}(k)=\frac{V(k)}{\epsilon(k)}
\end{equation}
where $\epsilon(k)$ is the dielectric permittivity given in terms 
of the static limit of the longitudinal part of gluon 
self-energy~\cite{schneider}
\begin{eqnarray}
\label{eqn4}
\epsilon(k)=\left(1+\frac{ \Pi_L (0,k,T)}{k^2}\right)\equiv
\left( 1+ \frac{M_D^2}{k^2} \right) \quad , 
\end{eqnarray} 
The result for the static limit of the dielectric permittivity is the
perturbative one. If one assumes that huge non-perturbative effects
(like the string tension) survive above $T_c$, the same could be true also
for such a dielectric function. So, there is a {\em caveat} that this (linear) 
relation of dielectric function ($\epsilon$) on $M^2_D$ may 
also pick up modifications due to the presence of non-perturbative
effects above the deconfinement point. To get rid of the complexity
of the problem, we put all the non-perturbative effects (including the 
non-zero string tension) together in the effective charge ($2\sigma/m_D^2$)
of the medium modified potential which further depends of the Debye mass. 
The quantity  $V(k)$, the Fourier transform 
(FT) of the Cornell potential reads~\cite{chandra}:
\begin{equation}
\label{eqn5}
{\bf V}(k)=-\sqrt(2/\pi) \frac{\alpha}{k^2}-\frac{4\sigma}{\sqrt{2}\pi k^4}.
\end{equation}
Substituting Eq.(\ref{eqn4}) and Eq.(\ref{eqn5}) into Eq.(\ref{eq3})
and evaluation of the inverse Fourier-Transform of the RHS of
Eq.(\ref{eq3}) one obtains the r-dependence of the medium modified 
potential. The expression thus reads
\begin{eqnarray}
\label{vrt}
{\bf V}(r,T)&=&(\frac{2\sigma}{m^2_D}-\alpha)\frac{\exp{(-m_Dr)}}{r}\nonumber\\
&-&\frac{2\sigma}{m^2_Dr}+\frac{2\sigma}{m_D}-\alpha m_D
\end{eqnarray}
This potential has a long range Coulombic tail in addition to the 
standard Yukawa term. After taking the high temperature limit, the 
above potential takes the form:
\begin{eqnarray}
\label{lrp}
{V(r)}\sim -\frac{2\sigma}{m^2_Dr}-\alpha m_D
\end{eqnarray}
The above form (apart from a constant term) is a Coulombic type as in 
the hydrogen atom problem by identifying the fine structure constant 
$e^2$ with the effective charge $2 \sigma/m_D^2$. Since
the Debye mass $m_D$ always increases with the temperature, the effective
charge $2 \sigma/m_D^2$ gets waned as the temperature is increased. 
This makes the potential too shallow to bind $Q$, $\bar Q$.
This results the melting of the bound states. It is important to note 
here the difference between the screening of the charge and the 
screening of the range of the
potential. In the usual picture adopted to study the dissociation
of quarkonia through the Cornell potential, linear term vanishes above 
the critical temperature because string tension vanishes. So, above the 
critical temperature, the only nonvanishing term in the
potential is the attractive coulomb term which gets screened in a 
Yukawa form making the potential short-ranged. If the range of the
potential becomes too short compared to the Bohr radius
it will be dissolved into its constituents. However, in our
case, this dissociation happens due to screening of the
charge, not due to screening of the range of potential.
Note that the constant terms in the potential (Eq.\ref{vrt})
are needed in computing the masses of the quarkonium states.
It is equally important while comparing our effective potential 
(Eq.\ref{vrt}) with the free energy in lattice studies (discussed
below). However, the constant terms are not needed while
comparing the values of the dissociation temperatures 
obtained in our model with the values in the lattice spectral studies.
This is due to the different criterion is imposed to evaluate the 
dissociation temperatures (discussed in the next section).

We need to mention that our in-medium effective potential $V(r,T)$ in 
Eq.(\ref{vrt}) agrees qualitatively (and also quantitatively) with the 
singlet part of the free energy in the lattice QCD~\cite{pot_1}. 
This have been checked by plotting $V(r,T)$ with $rT$ for a fixed value of 
$T/T_c$=3.32~\cite{vinet}.

We shall now systematically study the effects of perturbative 
and non-perturbative interactions on the binding energies and 
dissociation temperatures  of quarkonium states 
in a hot QCD medium by employing the three form of the
Debye mass(Eq.\ref{notation}).
In addition, we take advantage of all
the available lattice data, obtained not only in
quenched QCD ($N_f=0$), but also including two and, more recently, three light
flavors. We are then in a position to study also the flavor dependence of the
dissociation process, a perspective not yet achieved by the parallel studies of
the spectral functions, which are only available in
quenched QCD.\\

\noindent 
{\large{\bf Binding energy and dissociation temperatures}}

Binding energy of a quarkonium state at zero temperature 
is defined by the energy difference between the mass of the quarkonium 
and the open charm/bottom threshold. At finite temperature, the 
binding energy is defined as the distance between 
the peak position and the continuum threshold,
$E_{bin}=2 m_{c,b}+V_{\infty}(T) -M~$ with $M$ being the resonance 
mass~\cite{mocsyprl}. However, our definition is the conventional
one {\em viz.} the `ionization potential' in the atomic physics.
 
Finally, Schr\"{o}dinger equation for the potential
(Eq.\ref{lrp}) gives the energy eigen values for the
ground state and excited states, {\em viz.} $\jpsi$, 
$\psi^\prime$, $\Upsilon$, $\Upsilon^\prime$ etc. 
for the charmonium and bottomonium states, by the Bohr's 
formula:
\begin{eqnarray}
\label{bind1}
E_n=-\frac{E_I}{n^2} \quad; \quad E_I=\frac{m_q\sigma^2}{m^4_D}
\end{eqnarray}
Thus $E_n=-E_I, -\frac{E_I}{4}
-\frac{E_I}{9}, \cdot \cdot \cdot$ are the allowed energy levels
of $Q\bar{Q}$ bound states. These energies are known as ionization
potentials/binding energies for the $n$-th bound state. Thus, the binding 
energy becomes a temperature-dependent quantity through the 
Debye mass and it decreases with the temperature.
%%%%%%%%%%%%%%%%%%%%%%%%%%%%%%%%%%%%
\begin{figure*}
\vspace{-60mm}
\includegraphics[scale=.3]{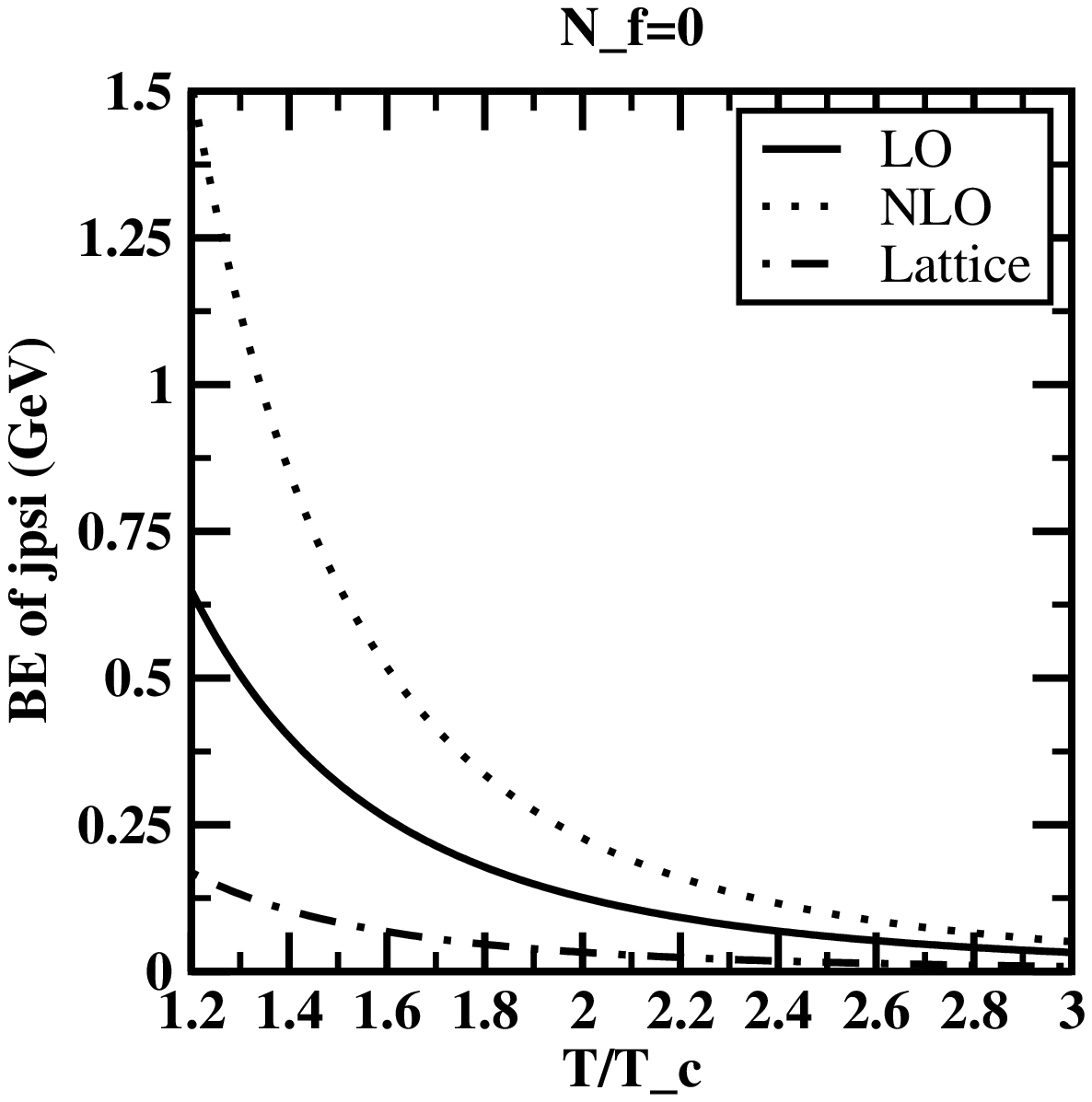} % Here is how to import EPS art
\hspace{-28mm}
\includegraphics[scale=.3]{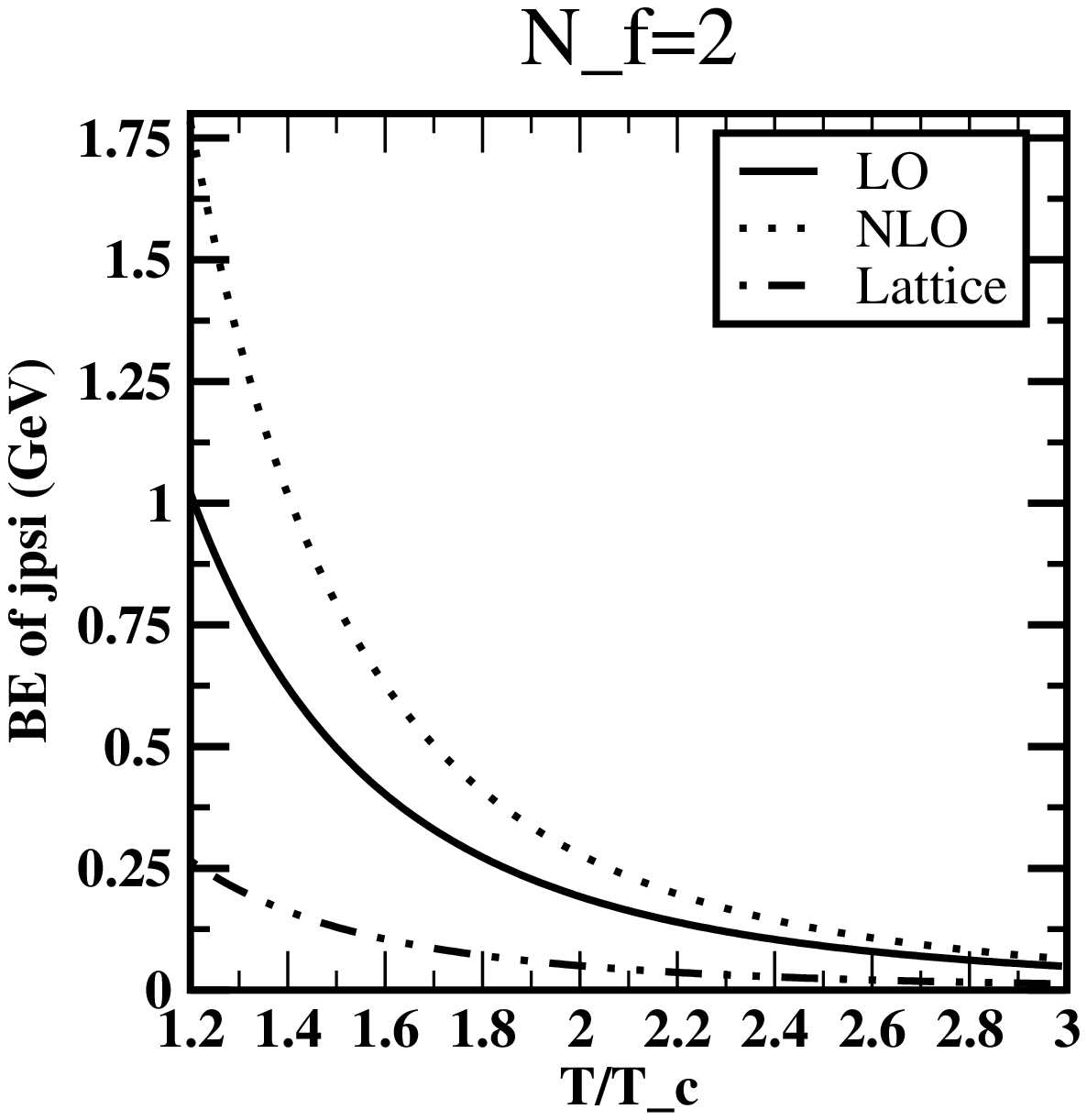}
\hspace{-28mm}
\includegraphics[scale=.3]{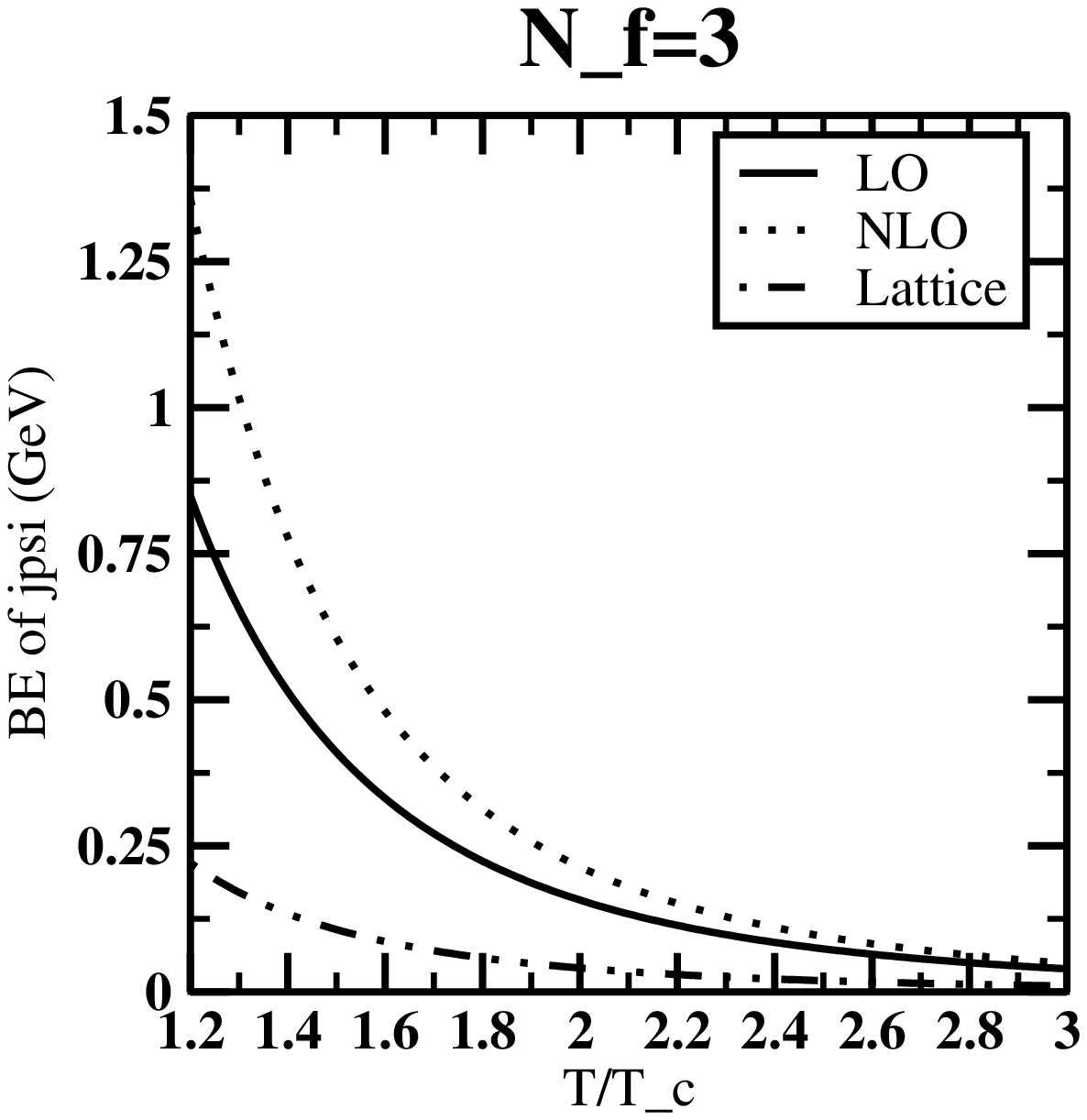}
\caption{Dependence of $J/\psi$ binding energy on temperature}
\end{figure*}

However, there are other states in the charmonium and bottomonium 
spectroscopy {\em viz} $\chi_c$'s and $\chi_b$'s and 
the binding energies for them are obtained from a variational
treatment of the relativistic two-fermion bound-state system in 
quantum electrodynamics~\cite{Dare1}.
\ba
\label{bind2}
E(\chi_{{}_{c,b}})= \frac{m_{{}_{c,b}}\sigma^2}{4 m^4_D} \left( 1 + \frac{2}{3}
 \frac{\sigma^2}{m^4_D} \right).
\ea
%%%%%%%%%%%%%%%%%%%%%%%%%%%%%%%%%%%%%%%%%%%%%%%%%
Figures 1-2 show the variation of binding energy with
temperature (in units of $T/T_c$) for the \J and \U states, 
respectively. Different curves 
in the figure denote the choice of Debye mass in Eq.(\ref{notation})
used to calculate the binding energy from 
Eq.(\ref{bind2}) or Eq.(\ref{bind1}). We consider three cases for 
our analysis: pure gluonic medium, 2-flavor
and 3-flavor to see the flavor dependence of dissociation
pattern in QCD medium.

The binding energy of the \J ($\approx$ 640 MeV) is
considerably larger than the typical non-perturbative hadronic
scale $\Lambda_{\rm{QCD}}$. As a consequence, perturbative term 
(leading-order term) in the Debye mass takes care
variation of binding energy with temperature. The same argument 
holds good for $\Upsilon$ also.
%%%%%%%%%%%%%%%%%%%%%%%%%%%%%%%%%%%%
\begin{figure*}
\vspace{-5mm}
\includegraphics[scale=.3]{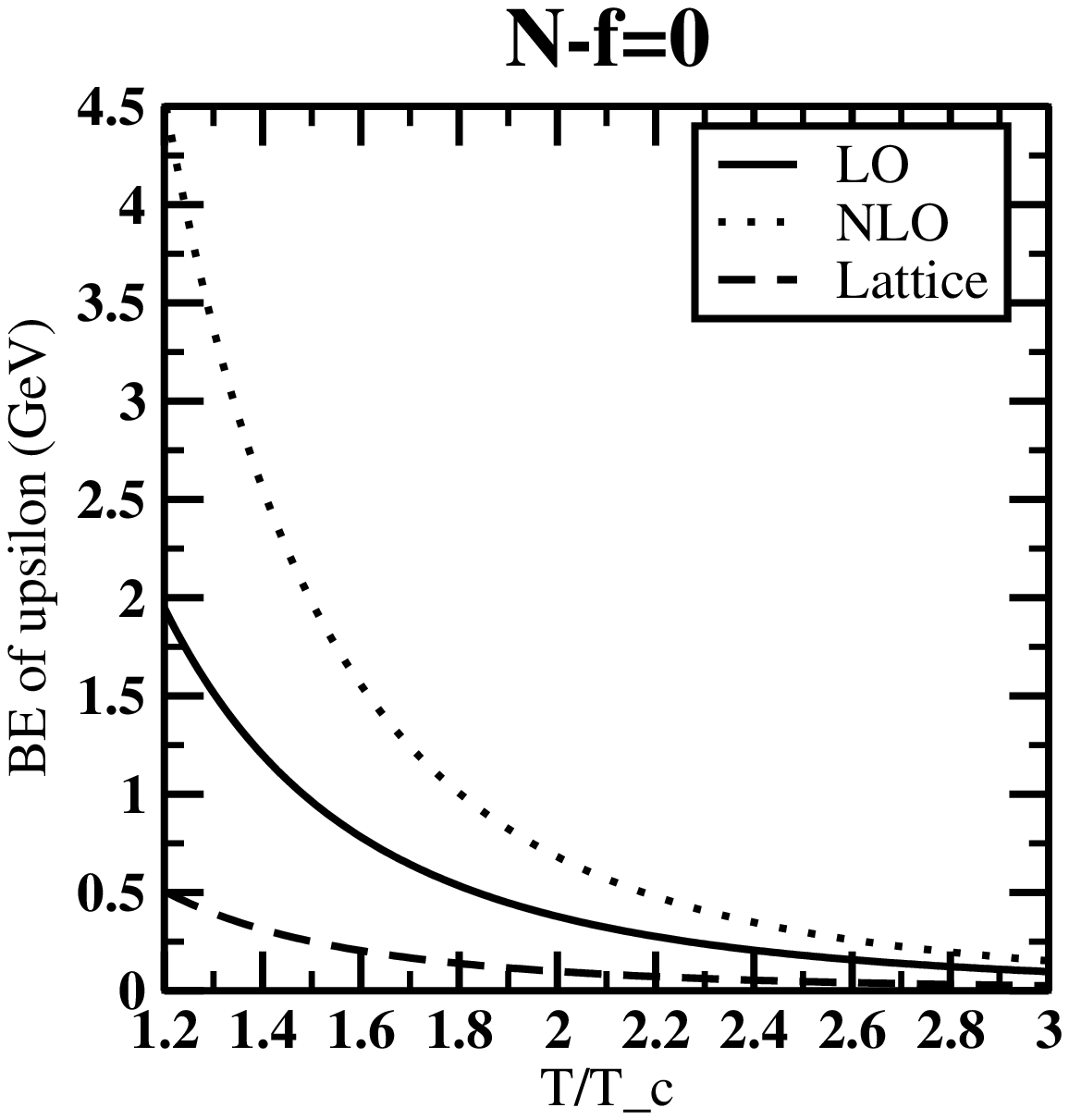}% Here is how to import EPS art
\hspace{-27mm}
\includegraphics[scale=.3]{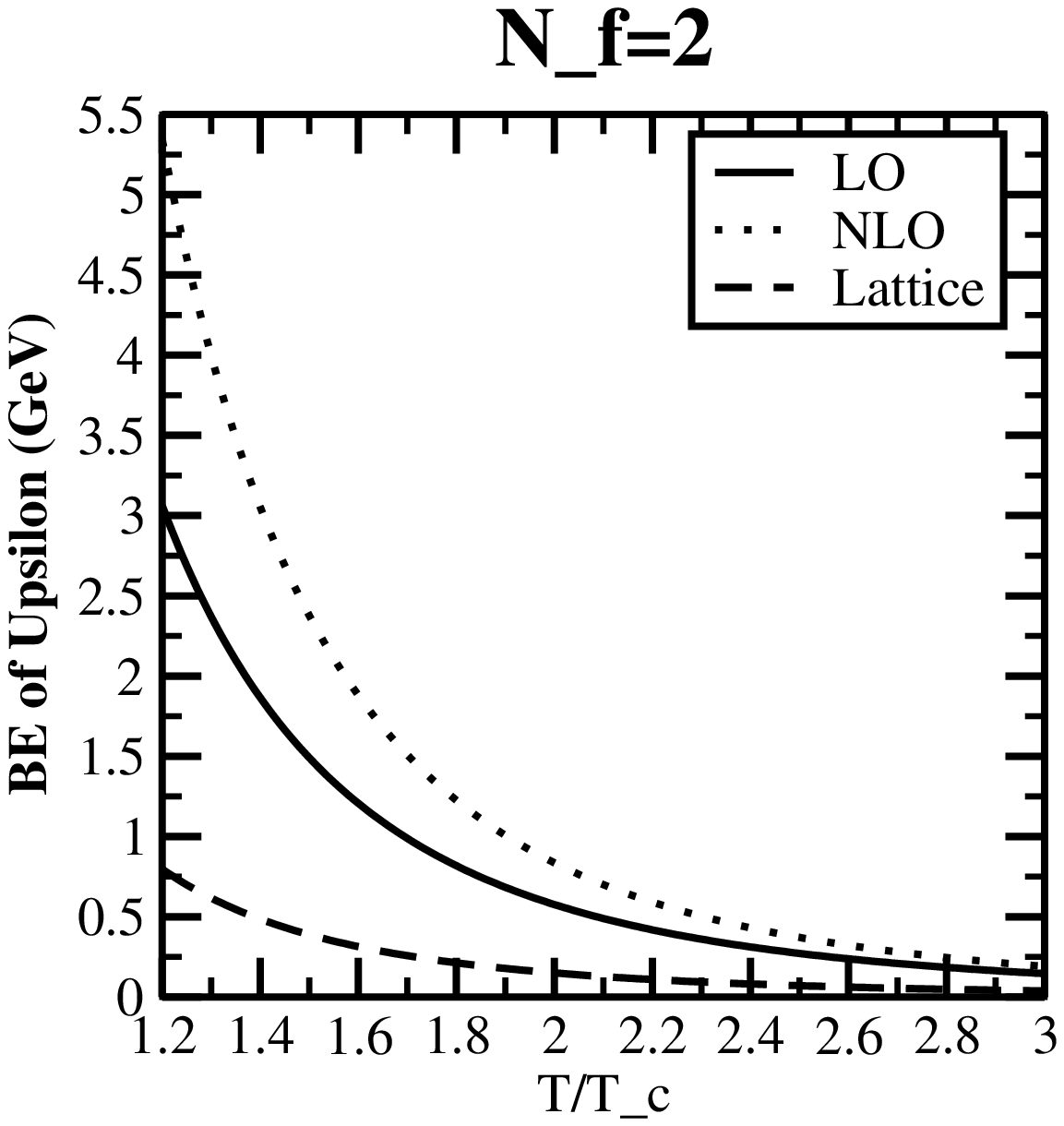}
\hspace{-28mm}
\includegraphics[scale=.3]{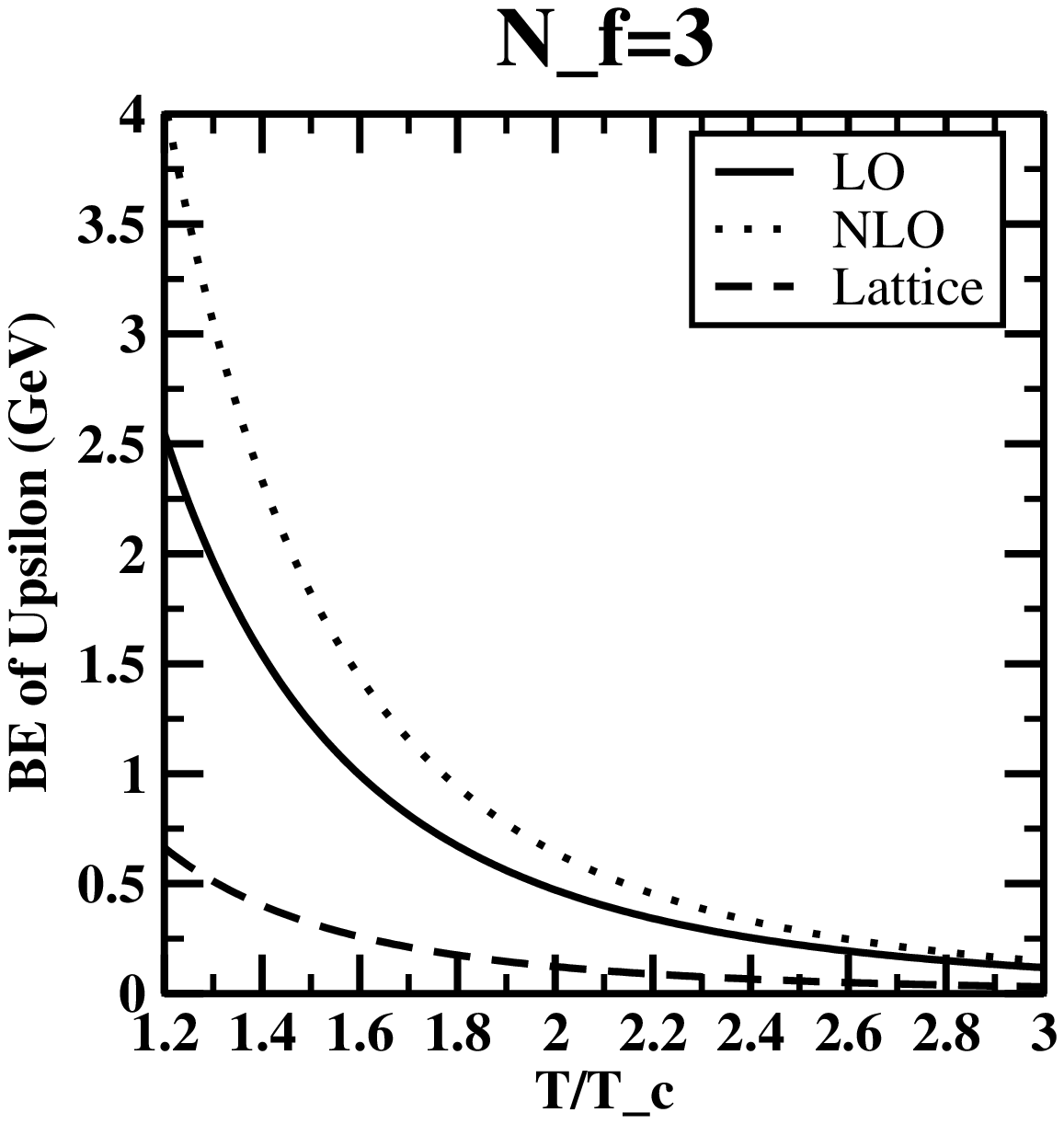}
\caption{Dependence of $\Upsilon$ binding energy on temperature }
\end{figure*}
%%%%%%%%%%%%%%%%%%%%%%%%%%%%%%%%%%%%%%%%%%%%%%%%%
Once we switch-on the non-perturbative contributions in the Debye mass
through the coefficients $c_{{}_{N}}$ and $d_{N,N_f}$, the
Debye mass becomes so large that the binding energies of all
the quarkonia becomes too small (even at  the temperature
100 MeV) compared to their ground state binding energies.
Indeed, the value of the Debye mass after inclusion of
nonperturbative terms is approximately three times larger than LO 
and NLO results and  twice as large as lattice parametrized Debye 
mass near $2T_c$. 
The temperature dependence of the binding energy for other
quarkonium states is studied in length in Ref.~\cite{vinet}.
However, this is not the complete story, the situation may change 
once the $O(g^4T)$ non-perturbative contributions to Debye mass are 
evaluated and utilize to estimate the binding energy for quarkonia states .

Thus the study of temperature dependence of binding energy will
help us to determine the dissociation temperatures of the quarkonium 
states in thermal medium.
Mocsy and Petreczky~\cite{mocsyprl} have defined the dissociation
temperature as the temperature above which the quarkonium spectral 
function shows no resonance-like structures, meaning that particular 
state is dissolved.

Physically, (thermal) dissociation of a bound state in a 
thermal medium can be explained as follows:
when the binding energy of a resonance ({\em viz.} $J/\psi$) state
drops below the mean thermal energy of parton
the state have become feebly bound and thermal
fluctuations can destroy it by transferring energy and exciting the
quark anti-quark pair into its continuum.
So, if the binding energy of a $c \bar c$ or $b \bar b$ state at some 
temperature becomes equal or smaller than the mean thermal energy 
then the state is said to be dissociated. 
Since the (relativistic) thermal energy of the 
partons is $3T$ hence the dissociation temperature $T_D$ of 
the $n$-th $Q\bar{Q}$ bound state will be determined by the
condition:
\begin{equation}
\label{bind1}
 \frac{1}{n^2} \frac{m_q\sigma^2}{m^4_D(T_D)}= 3T_D
\end{equation}
The above condition gives the dissociation temperatures after 
inserting expression for the Debye mass displayed in Eq.(\ref{notation}). 
However, the choice $3T$ is not rigid because even at low temperatures
$T < T_c$ (say) the Bose/Fermi distributions of partons will 
have a high energy tail with partons of mechanical energy 
$\epsilon > |E_n|$.

While determining the temperature dependence of the binding energy 
and dissociation temperatures the string tension is chosen to be 
$\sigma=0.184 {\rm{GeV}^2}$. The dissociation temperatures for 
the ground (1S), first excited states 
(2S), $\chi_c$, and $\chi_b$ (1P) of $c\bar{c}$ and $b\bar{b}$ are 
listed in the Tables 1 and 2 for the Debye masses in the
leading-order and the lattice parametrized form, respectively.
We do not put up the list for the dissociation temperatures
with the non-perturbative form of the Debye mass because
the values obtained are too small to explain physically.
We have taken the values of critical temperatures ($T_c$) 
270 MeV, 203 MeV and 197 MeV for pure gluonic, 
2-flavor and 3-flavor QCD medium, respectively~\cite{zantow}.

\begin{table}
\label{table1}
\caption{Dissociation temperatures for various quarkonia
(in unit of $T_c$) for $m^{LO}_D$.}
\centering
\begin{tabular}{|l|l|l|l|l|}
\hline
 Quarkonium state &Pure QCD & $N_f=2$&$N_f=3$\\
\hline\hline
$\jpsi$&1.1 &1.3 &1.2 \\
\hline
$\psi'$&0.8 &0.9 &0.9 \\
\hline
$\chi_c$&0.9 &1.1 &1.0 \\
\hline
$\Upsilon$&1.4 &1.7 &1.6 \\
\hline
 $\Upsilon'$&1.0 &1.2 &1.2 \\
\hline
$\chi_b$&1.1   &1.3  &1.2 \\
\hline
\end{tabular}
\end{table}
%%%%%%%%%%%%%%%%%%%%%%%%%%%%%%%%%%%%%%%%%%%%%%%%%%%%%%%%%%%%%5

\begin{table}
\label{table2}
\caption{Dissociation temperatures for various quarkonia
(in unit of $T_c$) for $m^{L}_D$.}
\centering
\begin{tabular}{|l|l|l|l|l|}
\hline
 Quarkonium state &Pure QCD & $N_f=2$&$N_f=3$\\
\hline\hline
$\jpsi$&0.8 &0.9 &0.9 \\
\hline
$\psi'$&0.5 &0.7 &0.6 \\
\hline
$\chi_c$&0.6 &0.7 &0.7 \\
\hline
$\Upsilon$&1.0 &1.2 &1.2 \\
\hline
 $\Upsilon'$&0.7 &0.9 &0.8 \\
\hline
$\chi_b$&0.7 &0.9 &0.9 \\
\hline
\end{tabular}
\end{table}

%%%%%%%%%%%%%%%%%%%%%%%%%%%%%%%%%%%%%%%%%%%%%%%%%%%%%%%%%%%%%5
The dissociation temperatures obtained
with the Debye mass in the leading-order (Table 1) is always 
larger than the dissociation temperatures with the Debye mass 
parametrized in lattice QCD ($m^{L}_D$) (Table 2).
for both charmonium and bottomonium states because of the larger value
of the Debye mass in lattice compared to LO values.
The results shown in the Tables 1-2 lend support with the recent 
lattice predictions~\cite{mocsyprl,satz}. The upper bound of the 
dissociation temperatures could be
obtained if average thermal energy is replaced by $\sim T$.\\

\noindent 
{\large {\bf Absorption by nucleons and comovers}}

So far we have discussed the fate of quarkonia only  when the
presence of quark gluon plasma is considered. It is very well
established that there are several aspects like initial state
scattering of the partons, shadowing of partons, absorption of
the pre-resonances ( $|Q\overline{Q}g>$ states) by the nucleons
before they evolve into physical quarkonia, and also dissociation
of the resonances by the comoving hadrons.
It has been argued that the absorption by co-moving hadrons
will be important for $\psi^\prime$, due to its very small binding
energy, while for more tightly bound resonances it may be
weak.

Let us briefly comment on them one-by-one. Shadowing of partons
should play an important role in the reduced production of quarkonia,
especially at the LHC energies. It is clear that if shadowing is
important, we shall witness a larger effect on $J/\psi$ than on $\Upsilon$,
because of the smaller values of the $x$ for gluons. At the same time, the
effect of shadowing should be similar for different resonances of the
charmonium (or bottomonium), as similar $x$ values would be involved for
them.

The absorption of the pre-resonances by the nucleons is another
source of $p_T$ dependence. It is important, to recall once
again that as the absorption is operating on the pre-resonance, the
effect should be identical for all the states of the quarkonium
which are formed.

This is a very important consideration as it is clear that if we
look  at the ratio of rates for different states of $J/\psi$ or
the $\Upsilon$ family as a function of $p_T$,
then in the absence of QGP-effects they would
be identical to what one would have expected in absence of nuclear
absorption and shadowing, providing a clear pedestal for the observation
of QGP.
%~\cite{ramona}.

Thes another aspect of $p_T$ dependence which needs to be
commented upon. The (initial state) scattering of
partons, before the gluons of the projectile and the target nucleons fuse
to produce the $Q\overline{Q}$-pair, leads to  an increase of the
$<p_T^2>$ of the resonance which emerges from the collision~\cite{shur}.
The increase in the $<p_T^2>$, compared to that for $pp$ collisions
is directly related to number of collisions the nucleons are
likely to undergo, before the gluonic fusion takes place. This leads to
a rich possibility of relating the average transverse momentum of the
quarkonium to the transverse energy deposited in the collision
(which decides the number of participants and hence the number of 
collisions).
Considering that collisions with large $E_T$ may have formation of QGP
in the dense part of the overlapping region, the quarkonia, which are
produced in the densest part (and hence contributing the largest
increase in the transverse momentum) are also most likely to melt
and disappear. This may lead to a characteristic saturation and even
turn-over of the $<p_T^2>$ when plotted against $E_T$ when the
QGP formation takes place. In absence of QGP, this curve would
continue to rise with $E_T$.\\

\noindent 
{\large {\bf Conclusions and Outlook}}

We have reviewed different theories/mechanisms of dissociation 
phenomena of heavy quarkonia in different stages of system produced in the 
relativistic nuclear collisions. Because the system formed 
just after the collision is not static and (thermally and chemically) 
equilibrated, it is rapidly expanding,
hadronize, and finally produces many pions, photons, leptons etc.
detected at the detectors. We discussed the dissociations 
mainly in three stages: initial state scattering/absorption, 
plasma interactions, hadronic/comover absorptions.
However, we gave much emphasis on the second stage of the
dissociation: dissociation in the deconfined medium/plasma. 
Apart from a comprehensive survey of different
approaches in lattice QCD and potential based studies, we
present our very recent work on the dissociation 
of quarkonia in a hot QCD medium by investigating the 
in-medium modifications to heavy 
quark potential. This is something new because, in our
formalism, medium modification results the (dynamical) screening
of the color charge in contrast to the screening of the range
of the potential in the usual screening picture.\\
The screening of the effective charge, in turn, causes the 
energy of the quarkonium state depends on temperature. 
We have then systematically studied the 
temperature dependence of the binding energy of the 
ground ($1S$) state, first excited ($2S$) state, and $1P$ states of 
charmonium and bottomonium in the pure and realistic QCD medium. 
We then determined the dissociation temperatures with the
Debye mass in leading-order $m_D^{\rm{LO}}$ and in the lattice 
parametrized form.  
The results are reasonably close to the finding of 
other theoretical works based on potential models~\cite{mocsyprl}. 
On the other hand these values are significantly smaller than 
the predictions made by others~\cite{satz,Alb07}.

In the end, we conclude that all the well explored and yet non-QGP effects 
need to be accounted for, before we can begin to see the suppression of the
quarkonium due to the formation of QGP. It seems that this has
been achieved at least at the SPS energies.\\

\end{document}